# Solid-state Janus nanoprecipitation enables amorphous-like heat conduction in crystalline Mg$_3$Sb$_2$-based thermoelectric materials

Rui Shu, Zhijia Han, Anna Elsukova, Yongbin Zhu, Peng Qin, Feng Jiang, Jun Lu, Per O. Å. Persson, Justinas Palisaitis, Arnaud le Febvrier, Wenqing Zhang, Oana Cojocaru-Mirédin, Yuan Yu*, Per Eklund* and Weishu Liu*

R. Shu, Z. Han, Y.B. Zhu, P. Qin, F. Jiang, W. S. Liu
Department of Materials Science and Engineering
Southern University of Science and Technology
Shenzhen 518055, China
E-mail: liuws@sustech.edu.cn

R. Shu, A. Elsukova, J. Lu, P. Å. Persson, J. Palisaitis, A.le Febvrier, P. Eklund
Thin Film Physics Division, Department of Physics Chemistry, and Biology (IFM), Linköping University, Linköping SE-581 83, Sweden
E-mail: per.eklund@liu.se

Prof. W. Q. Zhang
Department of Physics
Southern University of Science and Technology
Shenzhen 518055, China

O. Cojocaru-Mirédin, Y. Yu
I. Physikalisches Institut (IA)
RWTH Aachen University
Sommerfeldstraße14, 52074 Aachen, Germany
Email：yu@physik.rwth-aachen.de

W. S. Liu
Guangdong Provincial Key Laboratory of Functional Oxide Materials and Devices
Southern University of Science and Technology
Shenzhen 518055, Guangdong, China

**Solid-state precipitation can be used to tailor materials properties, ranging from ferromagnets and catalysts to mechanical strengthening and energy storage. Thermoelectric properties can be modified by precipitation to enhance phonon scattering while retaining charge-carrier transmission. Here, we uncover unconventional Janus-type nanoprecipitates in $Mg_3Sb_{1.5}Bi_{0.5}$ formed by side-by-side Bi- and Ge-rich appendages, in contrast to separate nanoprecipitate formation. These Janus nanoprecipitates result from local co-melting of Bi and Ge during sintering, enabling an amorphous-like lattice thermal conductivity. A precipitate size effect on phonon scattering is observed due to the balance between alloy-disorder and nanoprecipitate scattering. The thermoelectric figure-of-merit ZT reaches 0.6 near room temperature and 1.6 at 773 K. The Janus nanoprecipitation can be introduced into other materials and may act as a general property-tailoring mechanism.**

# 1. Introduction

Tailoring thermal conductivity is important for the design of a wide range of materials. Electronic components or high-temperature protective coatings require efficient heat dissipation. In contrast, thermoelectric materials require low thermal conductivity to achieve high conversion efficiency in waste-heat harvesting and environmentally friendly cooling[1]. The efficiency of thermoelectric materials is determined by the dimensionless figure of merit, defined as $ZT=S^2\sigma T/(\kappa_e+\kappa_{lat})$, where S, $\sigma$, T, $\kappa_e$ and $\kappa_{lat}$ are the Seebeck coefficient, electrical conductivity, absolute temperature, charge-carrier thermal conductivity and lattice thermal conductivity, respectively. The search for high-ZT materials has drastically progressed since the 1990s by introducing defects and/or reducing dimensionality, aiming at increasing the power factor ($S^2\sigma$) and decreasing $\kappa_{lat}$. Typical approaches include employing atomic-level disorder[2], dislocations[3], and nanoprecipitates[4] to introduce different scattering effects of the confined interfaces on carriers and phonons. In particular, introducing secondary-phase precipitates over multiple length scales is a general strategy for reducing heat transport by scattering phonons[5].

The "nanoparticle-in-alloy" concept has been demonstrated as an effective scattering mechanism for mid-to-low-frequency phonons[6]. In thermoelectric materials, these nanoparticles are often formed by either precipitation in a homogenous and supersaturated solid solution, or external inclusion during encapsulation. However, in most cases, only single-type isolated precipitates are obtained[5,7]. Recently, side-by-side co-precipitation evidenced by atom probe tomography (APT) has been shown to improve the mechanical properties of low-carbon steels[8]. The introduction of such co-precipitation with different sizes, mass, and shapes for each part of the co-precipitates, as well as complex interfaces with the matrix, could enhance the phonon scattering. Conventional nanoprecipitation phonon scattering strategies rely on controlling the number density of precipitates. Yet, theoretical modeling has shown that particle size is also of prominent importance. For example, an optimized silicide particle size of ~5 nm in SiGe[6], and ~10 nm ErAs in $In_{0.53}Ga_{0.47}As$[4] can maximize the phonon scattering by balancing the short- and long-wavelength limit of the scattering cross section[6]. However, controlling the optimized size distribution experimentally remains challenging.

Here, we choose $Mg_3(Sb,Bi)_2$-based materials for investigation. They are promising n-type thermoelectric compounds with performance comparable to the benchmark n-type $Bi_2Te_{3-x}Se_x$ but without scarce tellurium[9,10]. The thermoelectric properties of $Mg_3(Sb,Bi)_2$ can be improved by tuning carrier scattering through chalcogen[11,12] or transition metal doping[13,14], as well as by vacancy[15,16] and grain boundary engineering [17,18]. Near room temperature, the power factor of $Mg_3(Sb,Bi)_2$ has thus been increased to ~35 µW cm$^{-1}$ K$^{-2}$,[19], and ZT can be largely improved if eliminating the grain boundary resistance[20], which is promising for room-temperature thermoelectric applications[9,10,21–24]. Recent theoretical studies uncover that, compared to other Zintl compounds, the enhanced phonon-phonon scattering in $Mg_3Sb_2$-based materials due to soft Mg bonds[25], locally asymmetric vibrations[26], and high anharmonicity[27], causes an inherently low $\kappa_{lat}$. However, the thermal conductivity of $Mg_3(Sb,Bi)_2$ is still much higher than its amorphous limit[28] and thus has the potential to be further reduced through introducing additional phonon scattering mechanisms such as alloying and precipitates, in particular near room temperature. The lattice thermal conductivity ($\kappa_{lat}$) near room temperature is ~0.93 W m$^{-1}$ K$^{-1}$ for undoped $Mg_3(Sb,Bi)_2$ [29] and can be further reduced to ~0.7 W m$^{-1}$ K$^{-1}$ by doping Nb, Co or Y at the Mg-sublattice [13,14,30]. The reduction of lattice thermal conductivity by the formation of Bi-rich precipitates has also been reported[9,31]. Yet, the critical issues of precipitation kinetics and diversity remain unexplored. Engineering the density and composition of nanoprecipitates in $Mg_3(Sb,Bi)_2$ can therefore demonstrate a reduction in lattice thermal conductivity and a general mechanism for structural tailoring of nanoprecipitates.

## 2. Results and Discussion

## 2.1 Morphology of Janus nanoprecipitates.

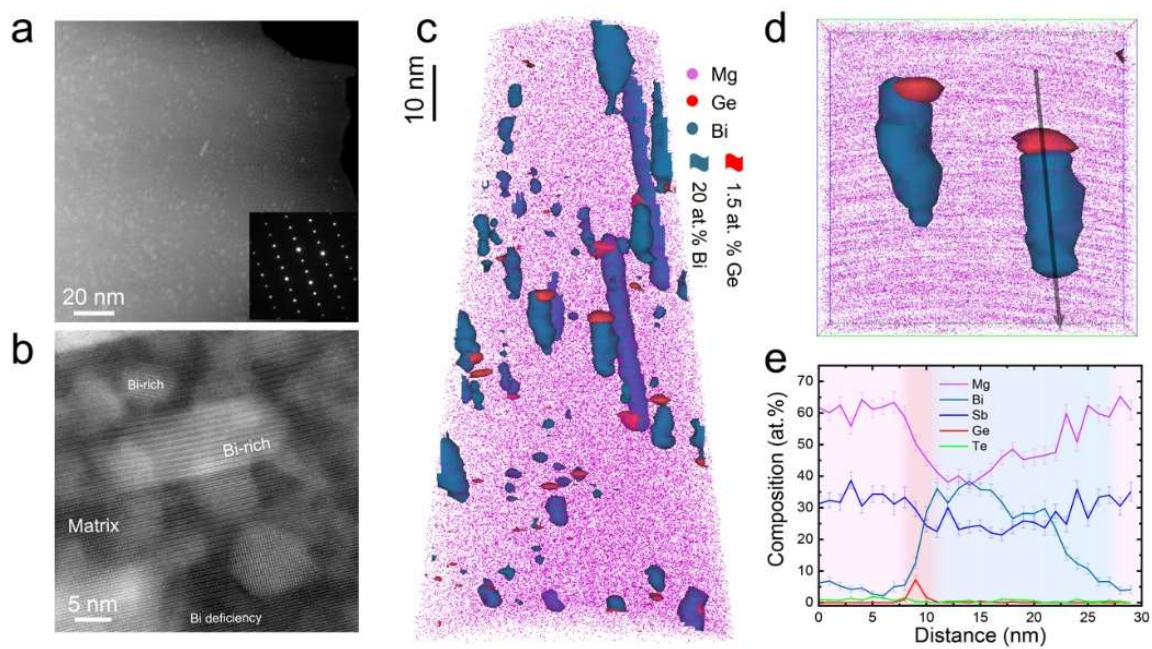

**Figure 1.** Microstructure of $Mg_{3.2}Sb_{1.47}Bi_{0.5}Te_{0.02}Ge_{0.01}$ specimen (Ge-0.01). a,b) STEM image, revealing a high density of nanoscale precipitates. The insets show the SAED pattern of the corresponding microstructure along the <100> zone axis. c) APT reconstructions showing the elemental distribution (Mg, pink; Ge, red; Bi, teal, Sb and Te atoms are omitted, for clarity). d) Close-up of a subregion from c, highlighting the 3D structure of Bi/Ge-rich Janus nanoprecipitates. The dimension of the cuboid region of interest is 30×30×10 nm$^3$. e) 1D composition profile along the arrow in **d**.

We have introduced high-density Bi/Ge-rich Janus nanoprecipitates in $Mg_{3.2}(Sb,Bi)_2$ through a local co-melting strategy during sintering. The $Mg_{3.2}(Sb,Bi)_2$ samples were doped with Te and Ge in the composition range from x = 0 to 0.05 in steps of 0.01, following the nominal formula of $Mg_{3.2}Sb_{1.49-2x}Bi_{0.5}Te_{0.01+x}Ge_x$, where Te acts as an effective n-type dopant to control the carrier concentration[11]. The structure of the $Mg_{3.2}Sb_{1.47}Bi_{0.5}Te_{0.02}Ge_{0.01}$ (Ge-0.01) specimen was investigated by scanning transmission electron microscopy (STEM) and APT. **Figure 1**a shows a high density of ultrafine precipitates (4-16 nm, brighter areas) uniformly distributed within grains. The selected area electron diffraction (SAED) pattern (inset in Figure 1a, <100> zone axis) exhibits an inverse α-$La_2O_3$ trigonal crystal structure of $Mg_3(Sb,Bi)_2$ without additional diffraction spots, indicating that the precipitates have the same crystal structure and coherent interfaces with the matrix. A magnified image (Figure 1b) reveals significant intensity fluctuations, which can be ascribed to compositional fluctuations in the $Mg_3Sb_{1.5}Bi_{0.5}$ matrix given that the high-angle annular dark field (HAADF) imaging contrast is proportional to $\sim Z^2$ (where Z is the atomic number). Bi-rich precipitates show bright contrast due to their large atomic number. Similar Bi-rich nanoprecipitates were also observed in Mn-doped $Mg_{3.2}Sb_{1.5}Bi_{0.5}$[9,31], but not in the Ge-free $Mg_{3.2}Sb_{1.5}Bi_{0.5}$ sample (Figure S1). This kind of local compositional fluctuations could result from the non-equilibrium synthesis route, i.e., ball milling and hot pressing. However, the number density of the Bi-rich nanoprecipitates in the as-fabricated Ge-doped $Mg_{3.2}Sb_{1.5}Bi_{0.5}$ is much higher than that in Mn-doped $Mg_{3.2}Sb_{1.5}Bi_{0.5}$[9]. To explore the fundamental mechanism, we conducted APT characterization, which constructs the 3D distribution of constituent elements with sub-nanometer spatial resolution and tens of parts-per-million elemental sensitivity[32]. Figure 1c shows an APT reconstruction of the Ge-0.01 sample, confirming the presence of a large volume fraction of Bi-rich precipitates highlighted by the iso-composition surface of 20 at.% Bi. The size of Bi-rich precipitates ranges between 2-20 nm in the APT reconstruction and is consistent with STEM observations (Figure 1d). The number density of Bi-rich precipitates is estimated to be $3.0 \times 10^{23}$ m$^{-3}$. Remarkably, a similarly high number of Ge-rich precipitates ($0.9 \times 10^{23}$ m$^{-3}$), as depicted by the iso-composition surface of 1.5 at.% Ge, are found to connect side-by-side with the Bi-rich precipitates, forming Janus particles, i.e., particles composed of two different phases on either side[33]. A close-up of the 3D morphology of Ge/Bi-rich Janus precipitates is shown in Figure 1d. The composition profile across the Janus particle in Figure 1e as indicated by the arrow in Figure 1d reveals an average composition of $28.6 \pm 1.5$ at.% Bi for Bi-rich precipitates, and $8.5 \pm 1.5$ at.% Ge for Ge-rich precipitates, which are much higher than the corresponding values of $9.5 \pm 1.5$ at.% Bi and $1.5 \pm 1.0$ at.% Ge in the matrix.

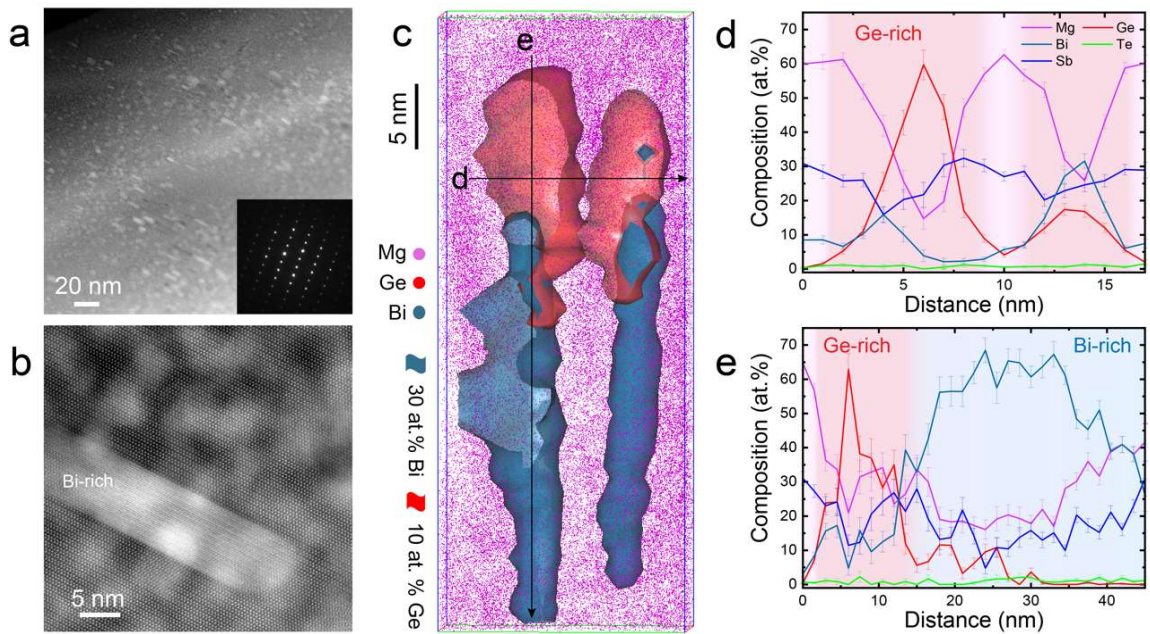

**Figure 2.** Microstructure of the $Mg_{3.2}Sb_{1.39}Bi_{0.5}Te_{0.06}Ge_{0.05}$ specimen (Ge-0.05). a) STEM image, revealing high-density nano-precipitates. The insets show the corresponding SAED pattern. b) Magnified view of a Bi-rich precipitate. c) APT reconstructions showing the elemental distribution (Mg, pink; Ge, red; Bi, teal); the Ge-rich and Bi-rich precipitates are depicted by iso-composition surfaces of 10 at.% Ge and 30 at.% Bi, respectively. Both precipitates connect side-by-side forming Janus precipitates. d) 1D composition profile calculated along the horizontal arrow across two Ge-rich precipitates. e) 1D composition profile calculated along the vertical arrow across the Janus precipitate from Ge-rich to Bi-rich part.

We further investigated the corresponding microstructure of $Mg_{3.2}Sb_{1.39}Bi_{0.5}Te_{0.06}Ge_{0.05}$ (Ge-0.05). The Janus nanoprecipitates also exist in a large volume fraction but are coarser than that in the sample with lower Ge contents. STEM images in **Figure 2**a and b show elongated nanoprecipitates in size of 5 - 50 nm embedded in the matrix (inset in Figure 2a, <100> zone axis, and XRD, see Figure S2). Furthermore, Figure 2c shows two Bi/Ge rich Janus nanoprecipitates with a size of ~50 nm in length and ~5 nm in diameter depicted by the iso-composition surface of 30 at.% Bi and 10 at.% Ge. The side-by-side configuration consolidates the unique Janus feature in the as-fabricated Ge-doped $Mg_{3.2}Sb_{1.5}Bi_{0.5}$. Note that two types of Bi-rich precipitates, i.e., Bi-rich $Mg_3(Bi,Sb)_2$ phase and pure Bi phase, are observed in the high Ge doping sample, which is confirmed by the STEM phase identification (Figure S3 and S4). The Ge-rich particle has been observed by STEM after *in situ* annealing treatment at 573 K and is confirmed with the same structure as the matrix (Figure 2a, Figure S5 and S6), i.e., $Mg_3(Bi,Sb)_2$ phase, but with high Ge content to 9 at.%.

Both Bi- and Ge-rich parts of the Janus nanoprecipitates coarsen with increasing Ge content. Based on the APT analysis, the number density of precipitates decreases from $3.0 \times 10^{23}$ m$^{-3}$ (Ge-0.01 sample) to $6.4 \times 10^{22}$ m$^{-3}$ (Ge-0.05 sample). The Ge content gradually decreases from the core of precipitates to the matrix, as revealed by the 1D composition profile across the Ge-rich precipitates (Figure 2d). The maximum composition of about 60 at.% Ge in the precipitate core is higher than that for the lower Ge-content sample (Figure 1e) due to the coarsening of precipitates with increasing Ge content. The content of Bi in the Bi-rich precipitates is also increased as determined by the 1D composition profile across the Janus particle from the Ge-rich part to the Bi-rich part (Figure 2e). Furthermore, in addition to the Ge-Bi configuration, we observed Ge-Bi-Ge precipitates, especially in samples with high Ge contents (Figure S7).

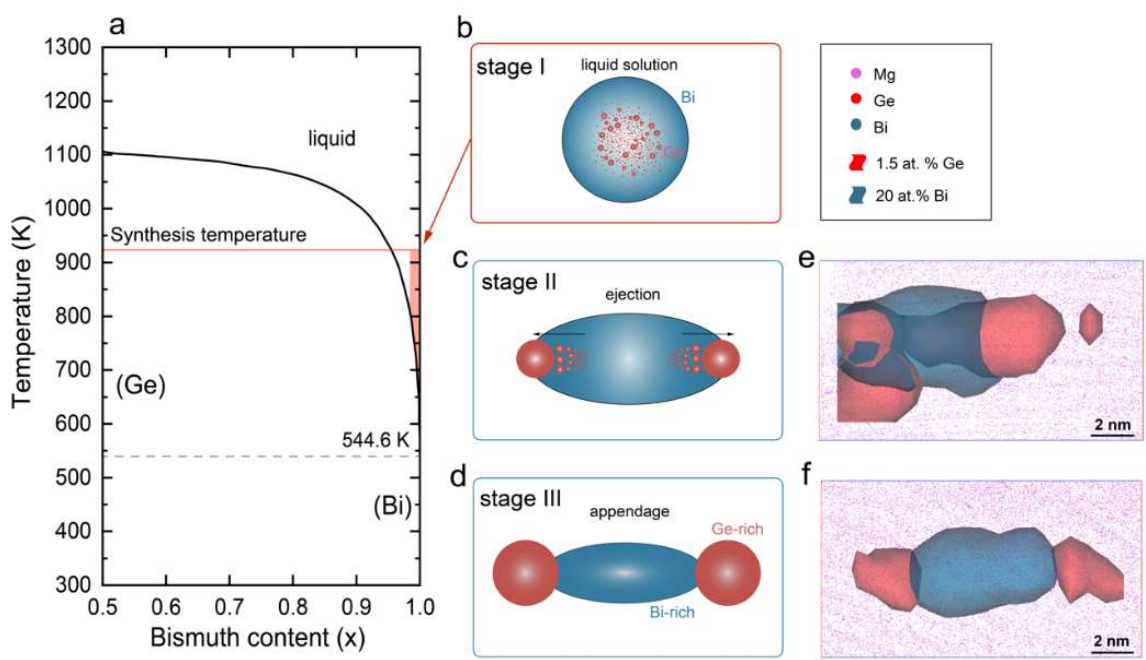

**Figure 3.** Mechanism of Janus precipitation. a) Bi-Ge binary phase diagram[34]. Red-colored areas denote the Ge-doping limitation for $Mg_{3.2}Sb_{1.5}Bi_{0.5}$ compounds. b,c,d) Schematic plots of the co-precipitation process during sintering and cooling, which can be divided into 3 stages. e,f) Experimental observation from APT reconstructions for correlative stage II, and III.

## 2.2 Kinetic mechanism of Janus nanoprecipitation.

The addition of Ge plays a critical role in the formation of Bi/Ge-rich Janus nanoprecipitation in the as-fabricated $Mg_{3.2}(Sb,Bi)_2$. **Figure 3**a shows the Bi-Ge binary phase diagram[34], with a colored area denoting the Ge-doping region in this work. The mutual solid-state solubilities of Ge and Bi are negligible, but Ge and Bi are fully miscible at the sintering temperature of 923 K. The formation of Bi/Ge-rich Janus nanoprecipitation could be explained by a co-melting mechanism. The as-fabricated $Mg_{3.2}(Sb,Bi)_2$ is made through a powder metallurgy route with high-energy ball milling (BM) followed by spark plasma sintering (SPS), which provides the energy necessary for the atomic diffusion into a crystalline structure at a temperature below the melting point of $Mg_3(Sb,Bi)_2$. Here, the sintering temperature is 923 K, higher than the melting point (544.6 K) of Bi. The formation of liquid Bi could be competitive with the crystallization of $Mg_3(Sb,Bi)_2$. At the first stage (Figure 3b), Bi forms a local liquid phase providing a molten reservoir for the Ge (melting point: 1211.5 K). The solubility of Ge in the solid Bi matrix is negligible, while it is 0.95 % in liquid Bi at 923 K. Without Ge, the local liquid Bi will be consumed by the crystallization of the $Mg_3(Sb,Bi)_2$ from the amorphous-like ball-milled powders. The crystallization of $Mg_3(Sb,Bi)_2$ and Bi-Ge liquid phase compete with each other. The addition of Ge shifts the balance of the two processes, resulting in the local Bi-Ge liquid phase remaining until the cooling process, in which the local Bi-Ge liquid phase is quickly encapsulated in $Mg_3(Sb,Bi)_2$ crystalline phase. At the second stage (Figure 3c), corresponding to a cooling process, the solute Ge starts precipitating and is ejected to the sides of the local Bi-rich liquid phase region. The Ge atoms do not simply thicken the shell, but rather form Ge appendages attached to the sides of the Bi-rich part. It can be explained from the view of thermodynamics and kinetics. The free energy reduction of the Ge-rich phase heterogeneously nucleated on Bi-rich precipitate exceeds the energy penalty of a Bi-/Ge-rich interface, and the insoluble Ge atoms diffuse in Bi-rich precipitates along <*a*, *b*> direction much easier than *c* direction. At the third stage (Figure 3d), corresponding to cooling down to the melting point of Bi, the local Bi-rich liquid phase solidifies, finally leading to a high density of Janus precipitates. Residual microstructural evidence of these two stages was observed in the medium Ge-doped sample (Ge-0.03, Figure S7), as shown in the APT reconstructions in Figure 3e and 3f. Generally, the addition of Ge plays a vital role in preventing the liquid Bi from being consumed by the crystallization of $Mg_3(Sb,Bi)_2$, resulting in the final Bi/Ge-rich Janus nanoprecipitation. From the viewpoint of thermodynamics, the contribution of the configuration entropy makes the Bi-Ge liquid phase more stable than the pure Bi liquid. Higher Ge content leads to a larger volume of local Bi-Ge liquid phase and finally larger size of Bi/Ge-rich Janus nanoprecipitation in the bulk materials.

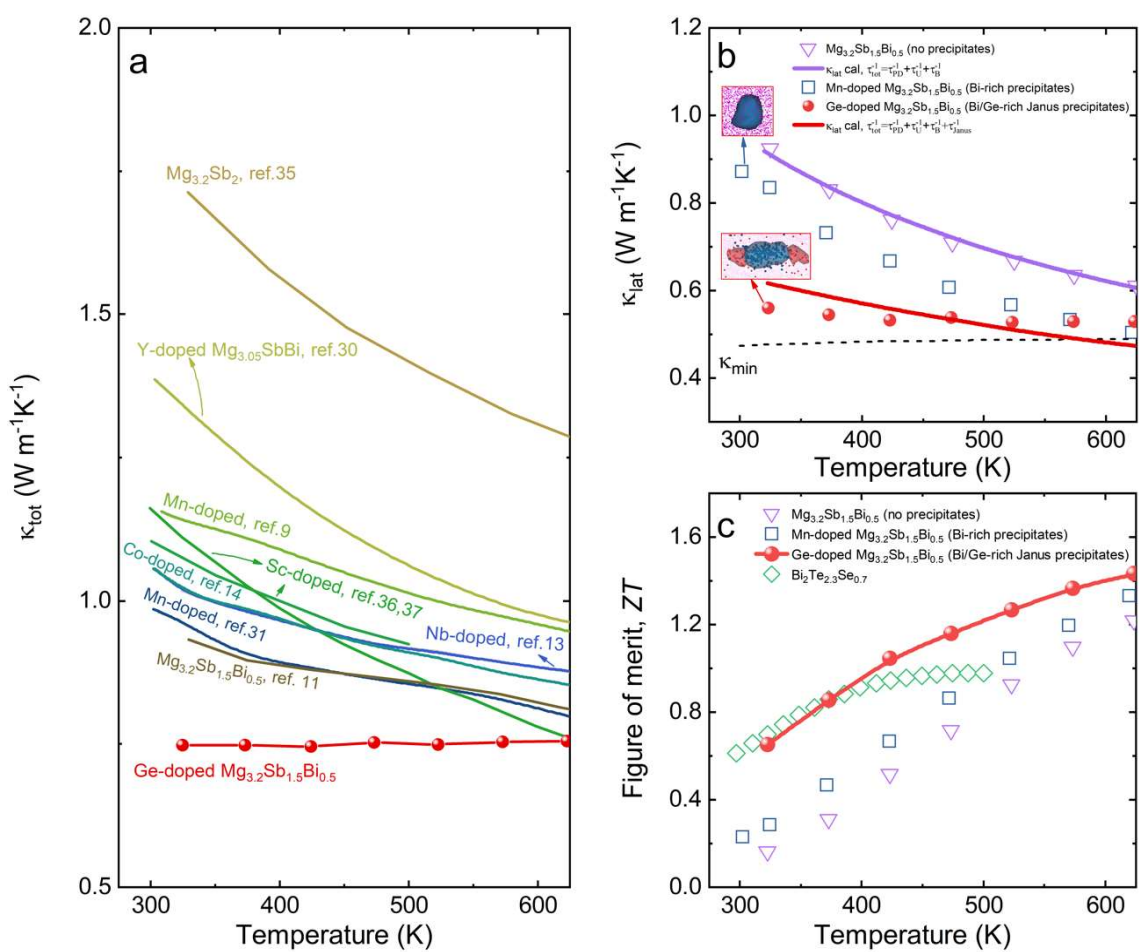

**Figure 4.** Thermoelectric performance of n-type Ge-doped $Mg_{3.2}Sb_{1.5}Bi_{0.5}$. a) A comparison of total thermal conductivity $\kappa_{tot}$ among several polycrystalline n-type $Mg_3(Sb,Bi)_2$ materials: $Mg_{3.2}Sb_2$[35], $Mg_{3.2}Sb_{1.5}Bi_{0.5}$ [11], transition-metal (Nb[13], Co[14], Mn[9,31], Sc[36,37], Y[30]) doped $Mg_3(Sb,Bi)_2$, and Janus precipitation $Mg_{3.2}Sb_{0.5}Bi_{0.5}$. b) Temperature dependence of lattice thermal conductivity $\kappa_{lat}$ of dopant-free $Mg_{3.2}Sb_{1.5}Bi_{0.5}$ (no precipitates), $Mg_{3.15}Mn_{0.05}Sb_{1.5}Bi_{0.5}$ [single-type Bi-rich precipitates, ref.[31]] and $Mg_{3.2}Sb_{1.47}Bi_{0.5}Te_{0.02}Ge_{0.01}$ (Bi/Ge-rich Janus precipitates). The solid purple and red lines in **b** show the temperature-dependent lattice thermal conductivity $\kappa_{lat}$

calculated with the Callaway model and input parameters obtained from STEM and APT investigation of Ge-free and Ge-0.01 doping samples. c) Figure of merit, ZT of above three $Mg_{3.2}Sb_{1.5}Bi_{0.5}$-based samples in comparison with state-of-the-art n-type thermoelectric material $Bi_2Te_{2.3}Se_{0.7}$[38].

## 2.3 Reduction of $\kappa_{lat}$ and enhancement of ZT

**Figure 4**a shows temperature-dependent thermal conductivity of as-fabricated Ge-doped $Mg_{3.2}Sb_{1.5}Bi_{0.5}$ with Janus precipitations, compared with reported n-type $Mg_3Sb_2$-based materials[9,11,13,14,30,31,35–37]. The as-fabricated $Mg_{3.2}Sb_{1.5}Bi_{0.5}$ with Janus precipitates shows strikingly low and amorphous-like thermal conductivity values. We compared the lattice thermal conductivity of the as-fabricated pristine $Mg_{3.2}Sb_{1.5}Bi_{0.5}$, $Mg_{3.2}Sb_{1.5}Bi_{0.5}$ with single Bi-rich precipitate induced by Mn dopant, and $Mg_{3.2}Sb_{1.5}Bi_{0.5}$ with Bi/Ge Janus nanoprecipitate in Figure 4b. The electrical transport properties of these three samples are shown in Figure S8). The lattice thermal conductivity ($\kappa_{lat}$) was deduced from the subtraction of the total thermal conductivity by the electronic thermal conductivity ($\kappa_e$) calculated from the Wiedemann–Franz law, $\kappa_e = L\sigma T$ with the temperature dependent Lorenz number $L$ derived from the Seebeck coefficient[39]. The $\kappa_{lat}$ of $Mg_{3.2}Sb_{1.5}Bi_{0.5}$ is significantly decreased due to the strong phonon scattering by the Bi/Ge Janus nanoprecipitates, especially in the temperature range of 300 - 625 K. It is well known that the temperature-dependent $\kappa_{lat}$ of a normal crystal with an acoustic-phonon-dominated scattering mechanism follows an inverse power law, i.e., $\kappa_{lat} \propto T^{-1}$. However, the temperature-dependent $\kappa_{lat}$ of the Janus nanoprecipitate systems strongly deviates from the relationship of $\kappa_{lat} \propto T^{-1}$, and approaches the amorphous limit $\kappa_{lat} \propto T^0$. A comparison of the exponent value (r) in the $\kappa_{lat} \propto T^r$ relation between the pristine $Mg_{3.2}Sb_{1.5}Bi_{0.5}$ and the Janus-nanostructured $Mg_{3.2}Sb_{1.5}Bi_{0.5}$ in Figure 4b shows that the Janus-nanostructured $Mg_{3.2}Sb_{1.5}Bi_{0.5}$ has an exponent value of r = –0.1, which approaches the amorphous limit of $Mg_{3.2}Sb_{1.5}Bi_{0.5}$ determined using the Cahill model[40] (see Supplementary Materials). Although high-density nanoprecipitates are demonstrated to lower the lattice thermal conductivity for many thermoelectric materials, such as PbTe, this amorphous-like behavior has not been achieved through regular precipitation.

It is well known that phonons have a wider distribution in the mean free path than that of electrons[41]. In a classic treatment of the phonon scattering of the nanoparticles, the relaxation time is only related to the particle size[42]. Mingo et al.[6] has theoretically shown a size-dependent $\kappa_{lat}$ in SiGe with silicide nanoparticles, suggesting an optimized particle size, but lacking direct observation. In this Janus-nanostructured $Mg_{3.2}Sb_{1.5}Bi_{0.5}$, the size of Bi/Ge Janus nanoprecipitates could be tuned by different Ge contents and a size of ~11 nm for Bi-rich precipitates can most efficiently scatter the phonons (Figure S9). Furthermore, the calculated lattice thermal conductivity is provided, based on the Callaway model[43] and parameters from the STEM and APT investigations (Supplementary Materials, Section 8). The phonon scattering of the Janus particles could be considered as a combination of an average effect of the whole particle and partial contributions of the internal sub-nanoparticle. Considering that the matrix is a heavily doped system with spherical nanoparticles, we only consider the Rayleigh limit ($\sigma_l$). The frequency-dependent phonon relaxation time of Janus nanoparticles ($\tau_{Jnp}$) can be derived as:

$$\frac{1}{\tau_{jnp}} = \frac{\upsilon}{3} \frac{f_{\langle jnp \rangle}}{R_{\langle jnp \rangle}} \left( \frac{D_{\langle jnp \rangle} - D_0}{D_0} \right)^2 \left( \frac{\omega}{\upsilon} R_{\langle jnp \rangle} \right)^4 + \sum_i \frac{1}{3(m\upsilon)^3} \frac{f_{i-np}}{R_{i-np}} \left( \frac{D_{i-np} - D_{\langle jnp \rangle}}{D_{\langle jnp \rangle}} \right)^2 \left( \omega R_{i-np} \right)^4 \quad (1)$$

where $f_{jnp}$, $R_{jnp}$, $D_{jnp}$ are the average volume fraction, radius, and density of the Janus nanoparticle, respectively, $D_0$ is the density of the matrix, $\upsilon$ is the average phonon speed, ω is the angular frequency, and m is a correction factor. The first term of the right side of Eq. (1) considers the Janus particle as an entirety, while the second term of the right side of Eq. (1) summarizes all the sub-particles within the Janus particle. In principle, the Janus particle provides a wide frequency range of phonon scattering. The calculated $\kappa_{lat}$ for Ge-free $Mg_{3.2}Sb_{1.5}Bi_{0.5}$ matches the experimental data well as a function of temperature, and that for $Mg_{3.2}Sb_{1.47}Bi_{0.5}Te_{0.02}Ge_{0.01}$ with Janus nanoprecipitates is reasonably close to experimental data but has a stronger temperature dependence. It indicates that the Janus precipitate structure increases the frequency range for phonon scattering, and hence significantly reduces $\kappa_{lat}$ near room temperature. The thermoelectric properties of $Mg_{3.2}(Sb,Bi)_2$ with different contents of Ge dopant were characterized. An increased thermoelectric figure-of-merit ZT = 0.6 near room temperature (Figure 4c) and 1.6 at 773 K are achieved, corresponding to an enhancement of 170 % at room temperature and 8% at 773 K, respectively, compared with the Ge-free sample (Figure S10).

## 3. Conclusion

In summary, we report a strategy to tune thermal conductivity by engineering the hierarchical structural aspects of nanoprecipitates, including composition, phase, and morphology. In the as-fabricated polycrystalline bulk Ge-doped $Mg_{3.2}Sb_{1.5}Bi_{0.5}$, a local co-melting strategy between Bi and Ge induces the formation of Janus nanoprecipitates. Two distinct nanoprecipitates with different masses provide a way to scatter a broader frequency range of phonons compared with individual precipitate structures, resulting in enhanced thermoelectric properties at low and intermediate temperatures for n-type $Mg_{3.2}Sb_{1.5}Bi_{0.5}$. This liquid-encapsulation-induced Janus precipitation approach is expected to be applicable to other thermoelectric materials and more generally to reduce thermal conductivity. We have provided a simple model to address the phonon scattering from both the average effect of the whole particle and partial contributions of internal sub-nanoparticles. This work thus provides a new perspective for tailoring hierarchical structures of nanoinclusions for low thermal conductivity and thermoelectric properties.

## 4. Experimental Section

*Materials synthesis*: High purity magnesium turnings (Mg, 99.8%; Alfa Aesar), antimony shots (Sb, 99.999%; 5N Plus), bismuth shots (Bi, 99.999%; 5N Plus), tellurium shots (Te, 99.999%; 5N Plus), and germanium powders (Ge, 99.99%; Alfa Aesar) were weighed according to the stoichiometric composition of $Mg_{3.2}Bi_{0.5}Sb_{1.49-2x}Te_{0.01+x}Ge_x$ (x=0.006, 0.01, 0.02, 0.03, 0.04, and 0.05). The extra Mg was used for compensating for the loss during synthesis. All the elements were mixed into a stainless-steel ball milling jar in a glove box under an argon atmosphere with an oxygen level <0.1 ppm. The materials were ball-milled for 10 hours. The ball-milled powders were then loaded into a graphite die with an inner diameter of 15 mm in a glove box. The graphite die with loaded powder was immediately hot pressed at 923 K for 10 mins. The sintering was done in a vacuum atmosphere at a pressure of 50 MPa. The thickness of hot-pressed disks was about 12 mm.

*Structural characterization:* X-ray diffraction (XRD) was carried out on a PANalytical X'Pert powder diffractometer with a Cu source ($\lambda_{K\alpha} \approx$ 1.5406 Å) operated at 45 kV/40 mA. Scanning transmission electron microscopy high angle annular dark field (STEM-HAADF) imaging and STEM energy-dispersive X-ray spectroscopy (STEM-EDX) analysis, with a Super-X EDX detector, were performed in the Linköping's monochromated, high-brightness, double-corrected FEI Titan[3], operated at 300 kV. The specimen for TEM examination was prepared by mechanical grinding followed by $Ar^+$-ion milling using a Gatan 691 Precision Ion Polishing Systems at liquid nitrogen temperature.

*Atom Probe Tomography measurement:* Needle-shaped APT specimens with an apex diameter of about 50 nm for samples $Mg_{3.2}Bi_{0.5}Sb_{1.49-2x}Te_{0.01+x}Ge_x$ were prepared using a dual-beam focused-ion beam (SEM/FIB) microscope (FEI Helios 650 NanoLab) equipped with a micromanipulator according to the standard lift-out method. The last step of the tip sharpening process utilized a low voltage and current (5 kV, 8 pA) $Ga^+$ ion beam to minimize Ga implantation in the sample (Ga content of the region analyzed was < 0.01 at%). APT experiments were conducted on a Cameca LEAP-4000X Si equipped with a picosecond UV laser (wavelength 355 nm). The specimen was maintained at 40 K and a-laser energy of 10 pJ was used at a pulse rate of 200 kHz with a target evaporation rate of 5 ions per 1000 laser pulses. The ion flight path was 160 mm. Ions were detected using a position-sensitive detector with a detection efficiency of ≈50%. This detection efficiency is the same for all ions evaporated. The data collected were 3D reconstructed and analyzed using the program IVAS v.3.8.0.

*Thermoelectric characterization:* All the samples were cut into about 2.5 mm × 3 mm × 14 mm pieces, which were coated with a thin-layer BN to protect instruments, to simultaneously measure the electrical resistivity and Seebeck coefficient under a low-pressure helium atmosphere from RT to 623 K (ZEM-3; ULVAC Riko). The thermal diffusivity $D$ was measured by the laser flash method (Netzsch LFA 467) and the thermal conductivity κ was calculated from κ = $dDC_p$, where density ($d$) of the samples was measured by the Archimedean method, the specific heat ($C_p$) was taken from the previous studies[9]. Electrical and thermal transport properties were both measured from the same as-pressed disk in directions perpendicular to the direction in which the pressure was applied to the samples during synthesis.

**Supporting Information**

Supporting Information is available from the Wiley Online Library or from the author.

**ACKNOWLEDGMENTS**

**Funding:** The work was supported financially by the National Key Research and Development Program of China under Grant No. 2018YFB0703600, National Natural Science Foundation of China under Grant No. 51872133, Guangdong Innovative and Entrepreneurial Research Team Program under Grant No. 2016ZT06G587, the Tencent Foundation through the XPLORER PRIZE, Guangdong Provincial Key Laboratory Program (2021B1212040001) from the Department of Science and Technology of Guangdong Province, the Swedish Government Strategic Research Area in Materials Science on Functional Materials at Linköping University (Faculty Grant SFO-Mat-LiU No. 2009 00971), the Knut and Alice Wallenberg foundation through the Wallenberg Academy Fellows program (KAW-2020.0196) and support for the Linköping Electron Microscopy Laboratory, and the Swedish Foundation for Strategic Research (SSF) support under the Research Infrastructure Fellow RIF 14-0074, and the German Science Foundation (DFG) within the project SFB 917. The authors acknowledge Dr. Yecheng Zhou from Sun Yat-Sen University and Prof. Keke Chang from Ningbo Institute of Industrial Technology, CAS for useful discussion.

**Author contributions:** W. L. and R. S. initiated the study. W.L., P. E., and Y.Y. supervised the research. R.S., Z. H, and P. Q. synthesized samples and performed part of microstructure characterization and thermoelectric property measurements. Y. Z. R. S. F. J. and W. L. performed the theoretical calculations of lattice thermal conductivity in discussion with W. Z. R. S. prepared TEM specimens. A. E., R. S., J. L. and J. P. performed the STEM analysis with contributions from P. O. Å. P., A. le F. and P. E. O. C. M. initiated the APT measurements and contributed to discussion and interpretations. Y. Y. performed APT experiments and data analysis. R. S., W. L., P. E. and Y. Y wrote the manuscript with contributions from the co-authors. All co-authors read, edited, and commented on successive version of the manuscript.

**Conflict of Interests**

The authors declare no competing interests.

**Data Availability Statement**

The data that support the findings of this study are available in the Supporting Information of this article, and from the corresponding author upon reasonable request.

**Keywords**

$Mg_3Sb_2$, Thermalelectrics, Low thermal conductivity, Janus nanoprecipitation, Atom Probe Tomography

# Supplementary Materials for

## Solid-state Janus nanoprecipitation enables amorphous-like heat conduction in crystalline Mg$_3$Sb$_2$-based thermoelectric materials

Rui Shu, Zhijia Han, Anna Elsukova, Yongbin Zhu, Peng Qin, Feng Jiang, Jun Lu, Per O. Å. Persson, Justinas Palisaitis, Arnaud le Febvrier, Wenqing Zhang, Oana Cojocaru-Mirédin, Yuan Yu*, Per Eklund* and Weishu Liu*

R. Shu, Z. Han, Y.B. Zhu, P. Qin, F. Jiang, W. S. Liu
Department of Materials Science and Engineering
Southern University of Science and Technology
Shenzhen 518055, China
E-mail: liuws@sustech.edu.cn

R. Shu, A. Elsukova, J. Lu, P. Å. Persson, J. Palisaitis, A.le Febvrier, P. Eklund
Thin Film Physics Division, Department of Physics Chemistry, and Biology (IFM), Linköping University, Linköping SE-581 83, Sweden
E-mail: per.eklund@liu.se

Prof. W. Q. Zhang
Department of Physics
Southern University of Science and Technology
Shenzhen 518055, China

O. Cojocaru-Mirédin, Y. Yu
I. Physikalisches Institut (IA)
RWTH Aachen University
Sommerfeldstraße14, 52074 Aachen, Germany
Email：yu@physik.rwth-aachen.de

W. S. Liu
Guangdong Provincial Key Laboratory of Functional Oxide Materials and Devices
Southern University of Science and Technology
Shenzhen 518055, Guangdong, China

**The amorphous limit of lattice thermal conductivity**

The amorphous limit of lattice thermal conductivity for $Mg_{3.2}Sb_{1.5}Bi_{0.5}$ was determined using the Cahill model [1]:

$$\kappa_{Lmin} = \left(\frac{\pi}{6}\right)^{1/3} k_B n^{2/3} \sum_{1} v_i \left\{\frac{T}{\theta_i}\right\} \int_0^{\theta_i/T} \frac{x^3 e^x}{(e^x - 1)^2} dx$$

where the sum is over the three sound modes with speeds of $v_i$, $n$ represents the number density of atoms, and the cutoff frequency ($\theta_i$) was determined as $\theta_i = v_i \left(\frac{h}{k_B}\right) \left(\frac{3n}{4\pi}\right)^{\frac{1}{3}}$.

The average phonon speed $v$ was calculated based on the equation $\frac{1}{v^3} = \frac{2}{3v_t^3} + \frac{1}{3v_l^3}$, where the transverse and longitudinal wave velocities, $v_t$ and $v_l$, were measured using a commercial resonant ultrasound spectroscopy (RUS) apparatus. The Lorenz number $L$ was calculated based on the single parabolic band (SPB) model and assuming that acoustic phonon scattering dominated.

# 1. Microstructure of Ge-free Mg$_{3.2}$Sb$_{1.5}$Bi$_{0.5}$

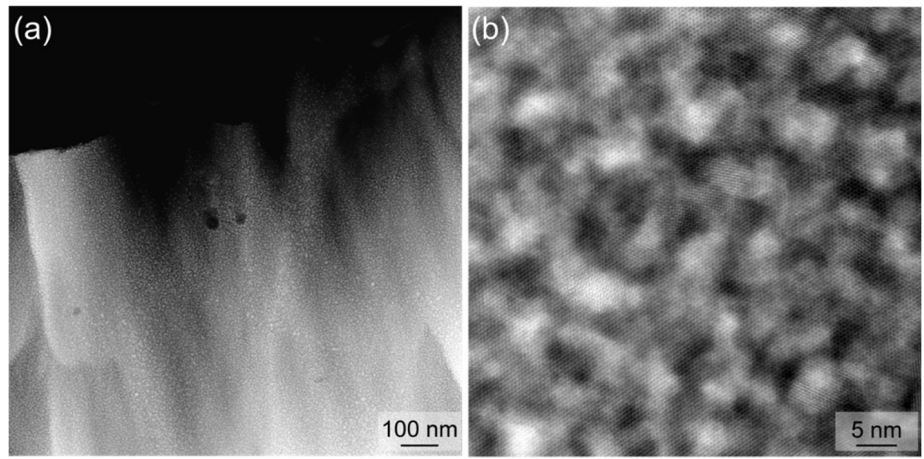

**Figure S1.** STEM images of Ge-free Mg$_{3.2}$Sb$_{1.5}$Bi$_{0.49}$Te$_{0.01}$ specimen. a) An overview STEM-HAADF image. b) HR-STEM HAADF image. Note that the bright and dark contrasts are due to composition fluctuations but not precipitates.

## 2. Phase composition and structure for n-type Mg$_{3.2}$Sb$_{1.5}$Bi$_{0.5}$ with different Ge-doping content

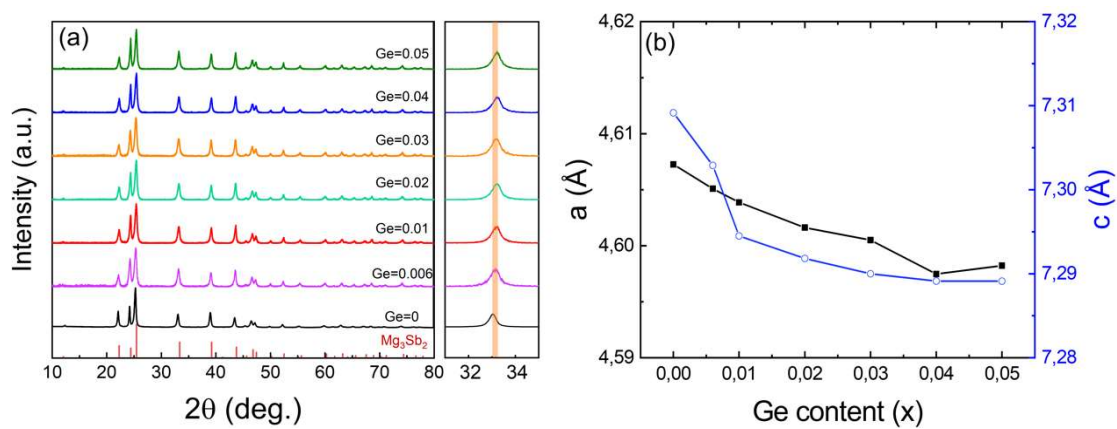

**Figure S2.** Structural characterization for n-type Mg$_{3.2}$Sb$_{1.5}$Bi$_{0.5}$ with different Ge-doping content. a) XRD patterns of as-fabricated Mg$_{3.2}$Bi$_{0.5}$Sb$_{1.49-2x}$Te$_{0.01+x}$Ge$_x$ specimens, no notable impurity phases were identified in this series of specimens. b) Lattice parameters of as-fabricated Mg$_{3.2}$Bi$_{0.5}$Sb$_{1.49-2x}$Te$_{0.01+x}$Ge$_x$ samples as a function of the Ge content x.

XRD patterns of as-fabricated Mg$_{3.2}$Bi$_{0.5}$Sb$_{1.49-2x}$Te$_{0.01+x}$Ge$_x$ samples with the nominal Ge compositions x =0, 0.006, 0.01, 0.02, 0.03, 0.04, 0.05 are displayed in Figure S2. All major reflections could be indexed to an inverse α-La$_2$O$_3$-type structure (space group, P$\bar{3}$m1) with no other impurity phase observed within the detection limits of powder XRD. The lattice parameters of Mg$_{3.2}$Sb$_{1.5}$Bi$_{0.5}$ solid solutions decreased with increasing Ge (x) and Te content. A trend is observed of decreasing lattice parameters calculated from the XRD patterns (Figure S2) and can be attributed to a reduction of the unit cell caused by Bi precipitates out from the Mg$_{3.2}$Sb$_{1.5}$Bi$_{0.5}$ solid solutions. This is consistent with the distance between (002) and (010) spots of Ge-0.05 sample in the corresponding SAED pattern (inset in Figure 2a in the main text) is shorter than that for the Ge-0.01 sample (Figure 1a in the main text).

## 3. Structure analysis of Bi- and Ge-rich precipitates and matrix observed in $Mg_{3.2}Sb_{1.39}Bi_{0.5}Te_{0.06}Ge_{0.05}$ specimen

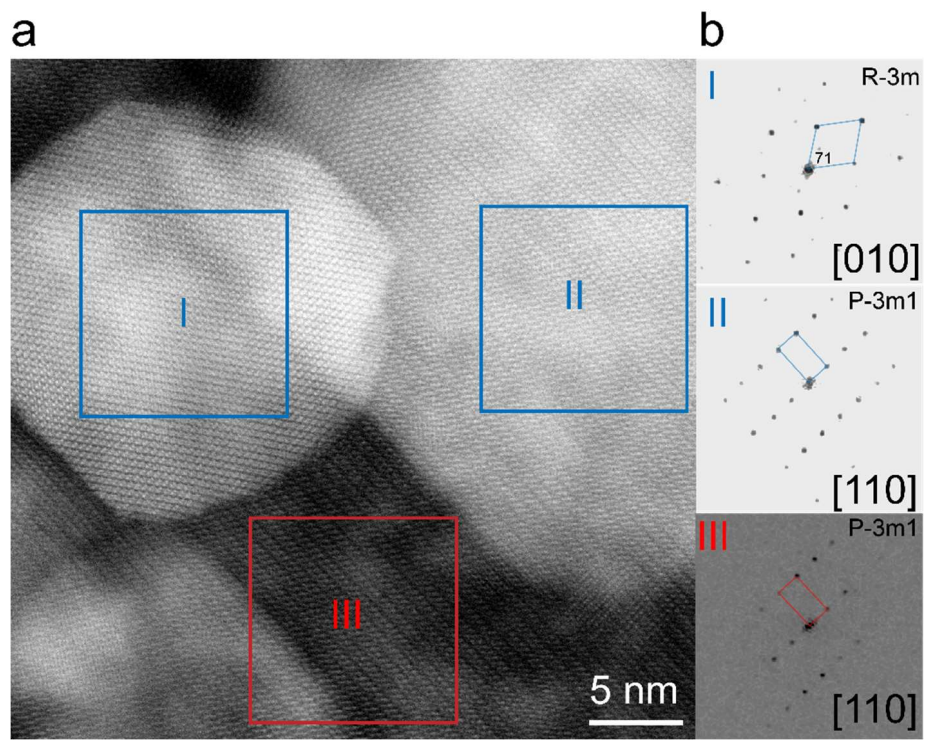

**Figure S3.** Identification of Bi-rich trigonal $Mg_3Bi_2$-type phase. a) HR-STEM-HAADF image of $Mg_{3.2}Sb_{1.39}Bi_{0.5}Te_{0.06}Ge_{0.05}$ specimen. b) FFT patterns of corresponding Bi-rich region I and II, and matrix region III.

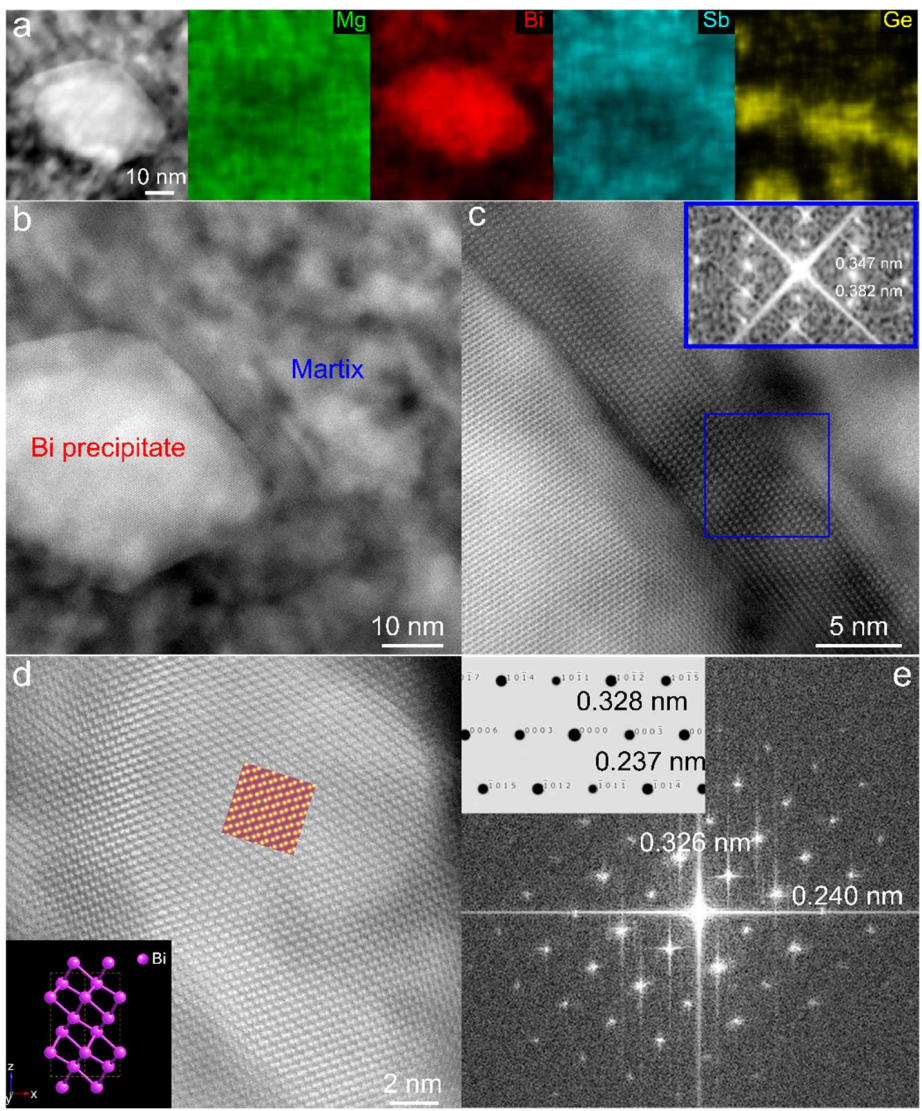

**Figure S4.** Structural characterization for Bi-rich precipitate of the $Mg_{3.2}Sb_{1.39}Bi_{0.5}Te_{0.06}Ge_{0.05}$ specimen. a) HAADF-STEM image of Bi-rich precipitate in $Mg_3Sb_{1.5}Bi_{0.5}$ matrix and corresponding Mg, Bi, Sb, and Ge elemental maps. b) HAADF-STEM image of Bi-rich precipitate and Mg-Sb-Bi matrix interface. c) High-resolution HAADF-STEM image of the precipitate-matrix interface. Inset: FFT pattern obtained from the region marked with blue square. d) High-resolution HAADF-STEM image of Bi-rich precipitate. Insets: crystal model of Bi (space group R-3m) viewed in [010] direction and corresponding simulated HAADF-STEM image. e) FFT pattern obtained from image d) Inset: Simulated diffraction pattern of Bi structure viewed in [010] direction.

We located Bi-rich precipitates by performing EDX elemental mapping (Figure S4a). Analysis of high resolution HAADF-STEM images of the precipitate (containing 61 at % of Bi) and its interface with matrix (Figure S4 b-e) showed that the precipitate has the structure of rhombohedral Bi.

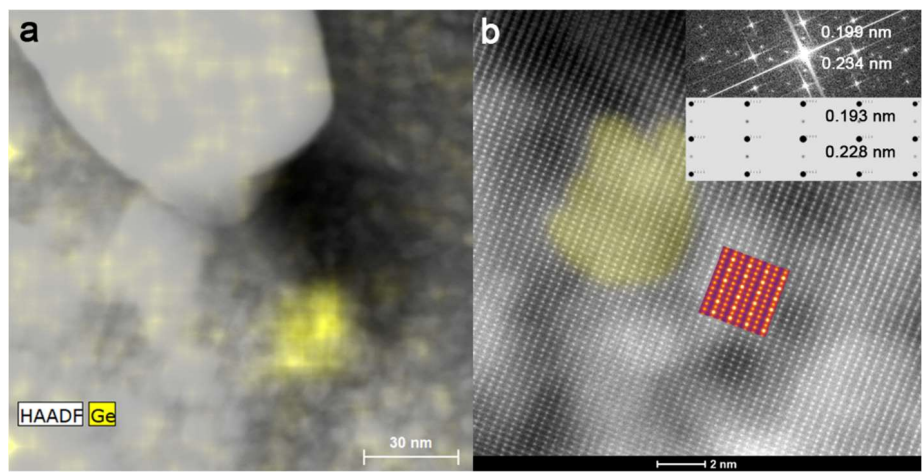

**Figure S5.** Structural characterization for Ge-rich precipitate of the $Mg_{3.2}Sb_{1.39}Bi_{0.5}Te_{0.06}Ge_{0.05}$ specimen. a) HAADF-STEM image with corresponding Ge elemental maps. b) High-resolution HAADF-STEM image of Ge-rich precipitate region. Inset on top, FFT pattern obtained from image b). Inset at the bottom, Simulated diffraction pattern of $Mg_3Sb_2$ (space group P-3m1) viewed in [120] direction and corresponding simulated HAADF-STEM image in the center.

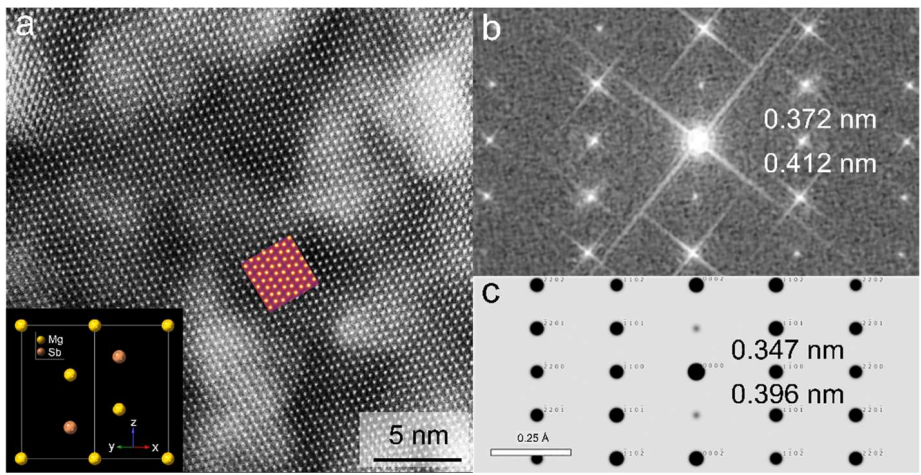

**Figure S6.** Structural characterization for the matrix of $Mg_{3.2}Sb_{1.39}Bi_{0.5}Te_{0.06}Ge_{0.05}$ specimen. a) High-resolution HAADF-STEM image of $Mg_3Sb_{1.5}Bi_{0.5}$ matrix. Lower inset: crystal model of $Mg_3Sb_2$ (space group P-3m1) viewed in [110] direction. Central inset: corresponding STEM simulation. b) FFT pattern obtained from image a. c) Simulated diffraction on $Mg_3Sb_2$ viewed in [110] direction.

The crystal structure of $Mg_3Sb_{1.5}Bi_{0.5}$ matrix for the $Mg_{3.2}Sb_{1.39}Bi_{0.5}Te_{0.06}Ge_{0.05}$ specimen was determined by comparing the contrast in HAADF-STEM image (Figure S6a) with STEM simulation (Figure 6a, central inset), and FFT pattern (Figure S6b) with the simulated diffraction pattern (Figure S6c) of the $Mg_3Sb_2$ P-3m1 structure viewed in [110] direction.

# 4. APT characterization on medium Ge-doping $Mg_{3.2}Bi_{0.5}Sb_{1.43}Te_{0.04}Ge_{0.03}$

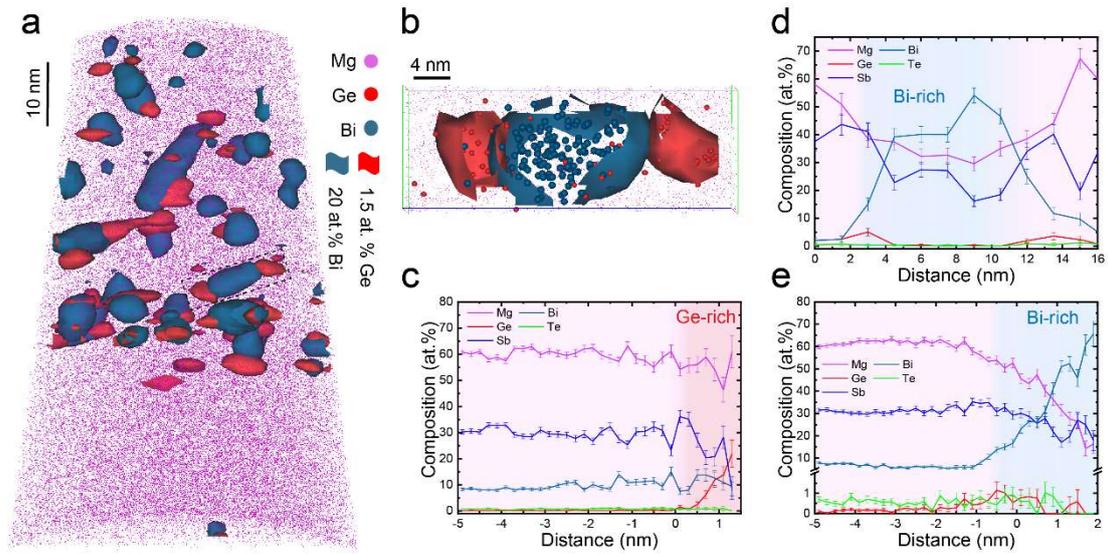

**Figure S7.** Three-dimensional elemental distribution analysis on nanostructures in the $Mg_{3.2}Bi_{0.5}Sb_{1.43}Te_{0.04}Ge_{0.03}$ specimen. a) APT reconstructions showing the distribution of elements (Mg, pink; Ge, red; Bi, teal); the Ge-rich and Bi-rich precipitates are depicted by iso-composition surfaces of 1.5 at.% Ge and 20 at.% Bi, respectively. Both precipitates connect side-by-side forming Janus precipitates. b) Close-up of a subregion taken from Figure S7a highlighting the 3D structure of Bi/Ge-rich Janus nanoprecipitates. c) composition proximity histogram calculated across the iso-composition surface of 1.5 at.% Ge showing the maximum Ge content at the precipitate core. d) 1D compositional profile along the horizontal direction of b showing the Bi-rich precipitate sandwiched by two Ge-rich precipitates. e) proximity histogram of 20 at.% Bi iso-surface showing the composition from the matrix to the Bi-rich precipitate core.

## 5. Comparison of thermoelectric performance

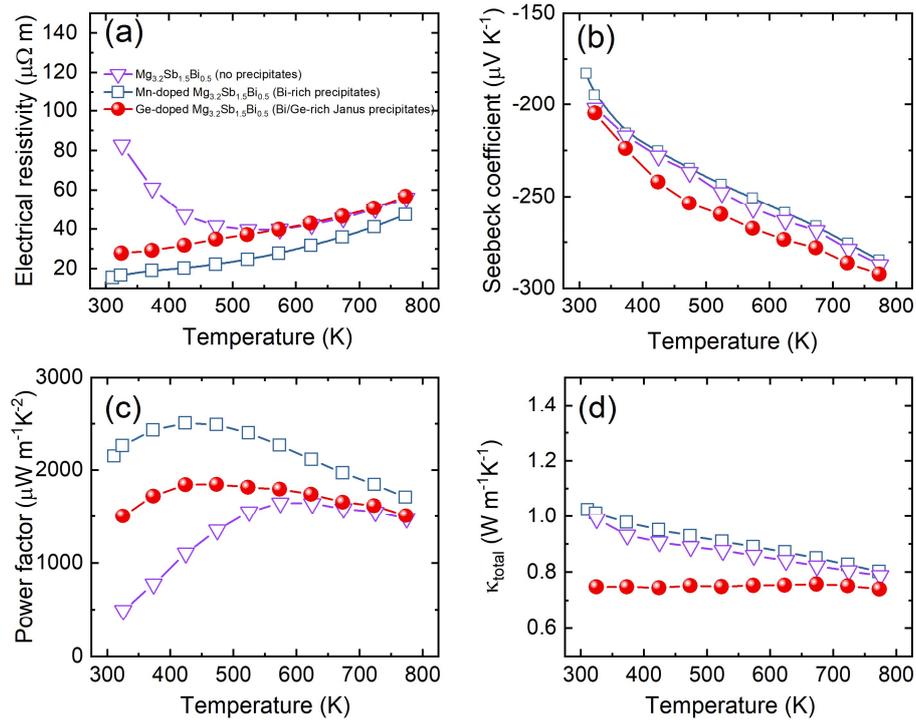

**Figure S8.** Temperature dependences of a) electrical resistivity, b) Seebeck coefficient, c) power factor, d) total thermal conductivity $\kappa_{tot}$, for dopant-free $Mg_{3.2}Sb_{1.5}Bi_{0.5}$ (no precipitates), $Mg_{3.15}Mn_{0.05}Sb_{1.5}Bi_{0.5}$ [single-type Bi-rich precipitates, ref.[2]] and $Mg_{3.2}Sb_{1.47}Bi_{0.5}Te_{0.02}Ge_{0.01}$ (Bi/Ge-rich Janus precipitates) in the direction perpendicular to the pressure applied to the samples during synthesis.

## 6. Precipitates-size effects on lattice thermal conductivity

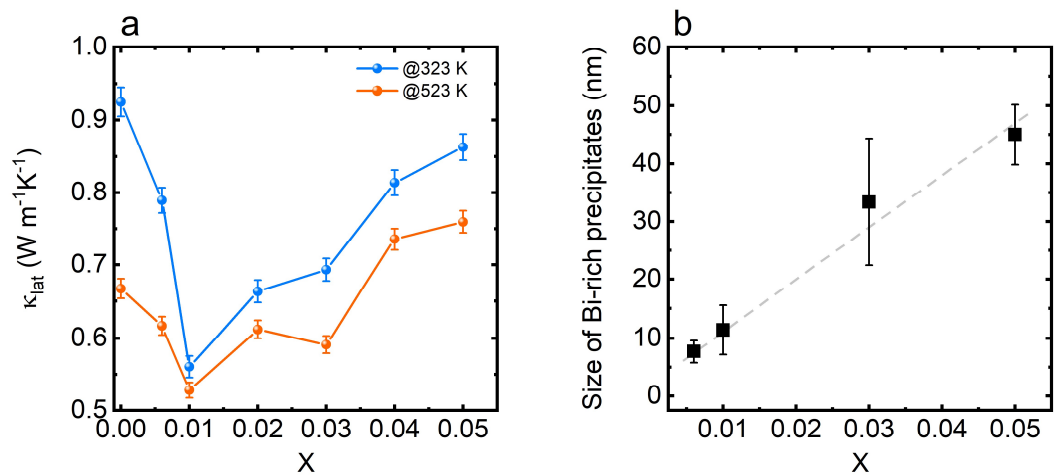

**Figure S9.** a) Lattice thermal conductivity $\kappa_{lat}$ at 323 K and 523 K as a function of Ge dopant fraction x. b) The size of Bi-rich precipitate determined from APT and HRTEM images as a function of Ge dopant fraction x.

## 7. Thermoelectric performance overview for Ge-doped Mg$_{3.2}$Sb$_{1.5}$Bi$_{0.5}$

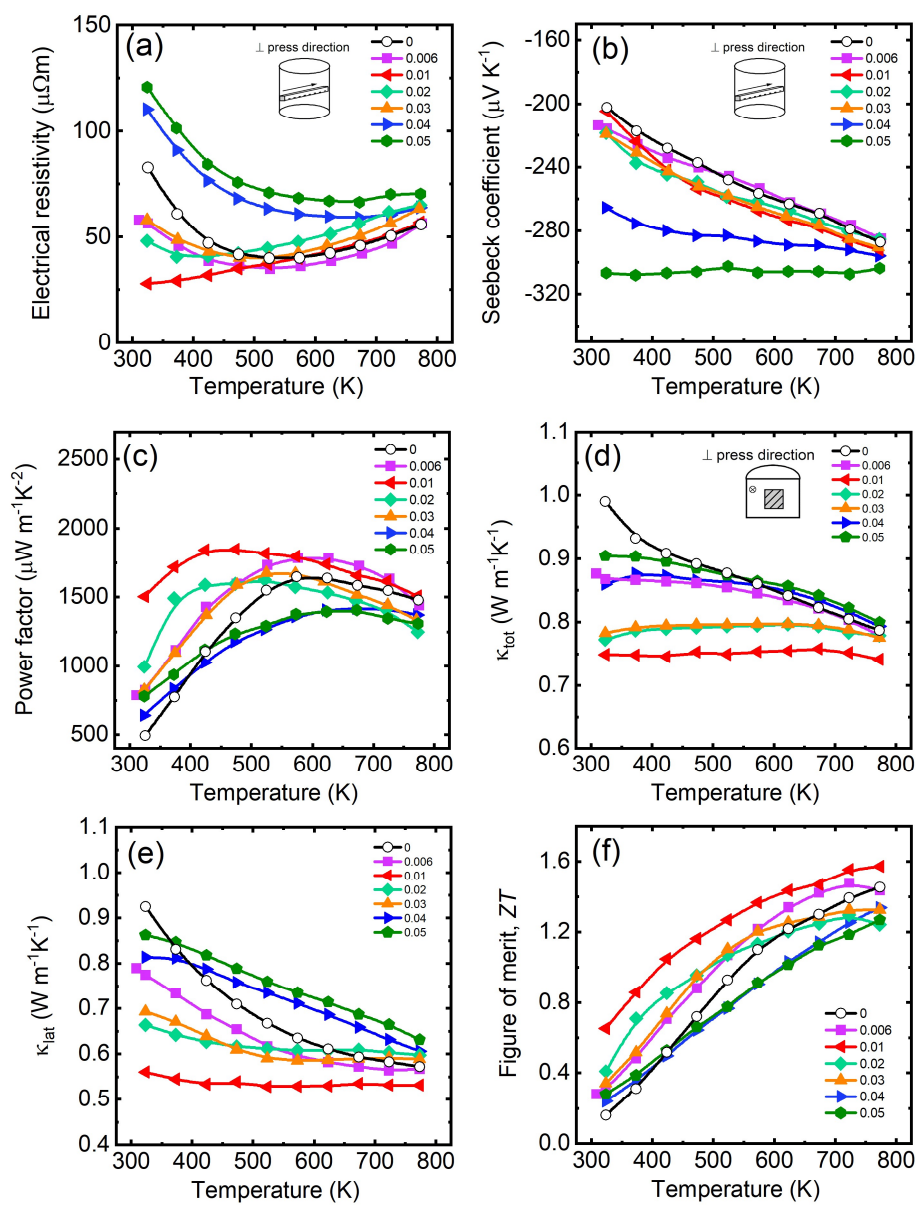

**Figure S10.** Electrical and thermal transport behavior in different dopant contents. Temperature dependences of a) electrical resistivity, b) Seebeck coefficient, c) power factor, d) total thermal conductivity $\kappa_{tot}$, e) lattice thermal conductivity $\kappa_{lat}$ and f) ZT values for Mg$_{3.2}$Sb$_{1.49-2x}$Bi$_{0.5}$Te$_{0.01+x}$Ge$_x$ specimens (x=0-0.05) in direction perpendicular ($\perp$) to SPS press direction.

The electrical resistivity of the x=0 sample first decreases and then increases with temperature. This is a typical phenomenon observed in polycrystalline Mg$_3$Sb$_2$-based compounds, which can be attributed to the grain boundary potential barrier scattering of electrons.[3] By adding a slight content of Ge (x=0.006 and 0.01), the barrier height decreases due to the segregation of Ge and Bi to grain boundaries. Yet, we found that the barrier height increases again with further increasing the content of Ge (Figure S10a), thus increasing the electrical resistivity, especially at low temperatures. Both STEM and APT proved that the size of nanoprecipitates increases with the Ge content. A larger particle can cause stronger charge carrier scattering at the interface and thus increase the electrical resistivity and the Seebeck coefficient due to potential energy filtering effect[4-6]. This explains the x=0.04 and x=0.05 samples showing the largest resistivity and absolute value of Seebeck coefficients. A suitable Ge content can compensate for the vacancy electron scattering in the pristine $Mg_{3.2}Sb_{1.5}Bi_{0.5}$ as reported in other transition-metal doped samples[5,6]. As a consequence, the x=0.01 sample shows the lowest electrical resistivity and the highest power factor (Figure S10). The extremely low thermal conductivity for the x=0.01 sample results from the phonon scattering of Janus particles and the size effect of nanoparticles, as discussed in the main text. Too large precipitates could not effectively scatter phonons due to the short phonon mean free path in $Mg_3Sb_2$. In contrast, the high intrinsic thermal conductivity of Bi and Ge could in turn increase the overall lattice thermal conductivity as observed in the high-content Ge samples. As a consequence, the optimized ZT values are obtained in the sample x=0.01. Note that the ZT values obtained in this work are not the highest reported in n-type $Mg_3Sb_2$ compounds. However, the Janus nanoprecipitation phenomenon observed in this work may be used to further enhance thermoelectric properties.

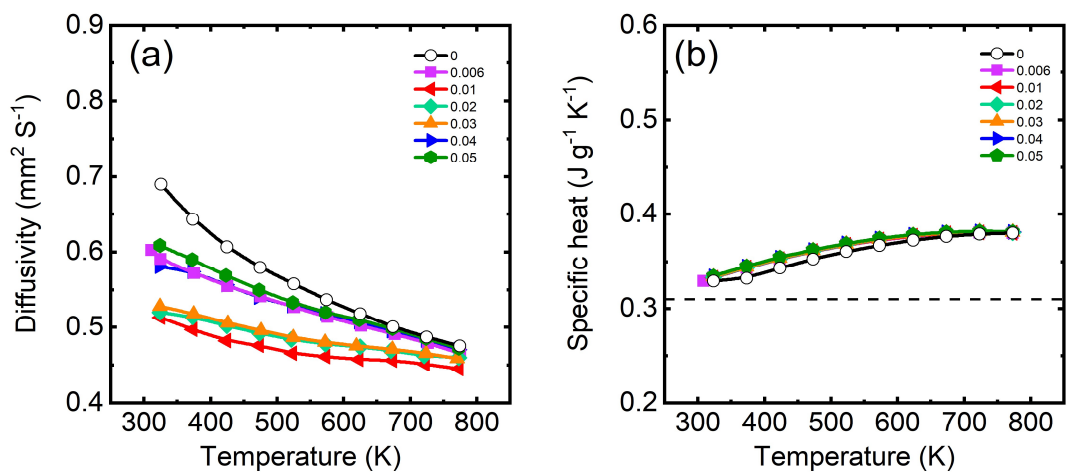

**Figure S11.** Thermal diffusivities and specific heat of $Mg_{3.2}Sb_{1.49-2x}Bi_{0.5}Te_{0.01+x}Ge_x$ specimens (x=0-0.05).

**Table S1.** Densities fof $Mg_{3.2}Sb_{1.49-2x}Bi_{0.5}Te_{0.01+x}Ge_x$ specimens (x=0-0.05).

| Specimen (x) | 0 | 0.006 | 0.01 | 0.02 | 0.03 | 0.04 | 0.05 |
|---|---|---|---|---|---|---|---|
| Density ($\pm 0.01$ g cm$^{-3}$) | 4.38 | 4.41 | 4.43 | 4.45 | 4.44 | 4.42 | 4.44 |

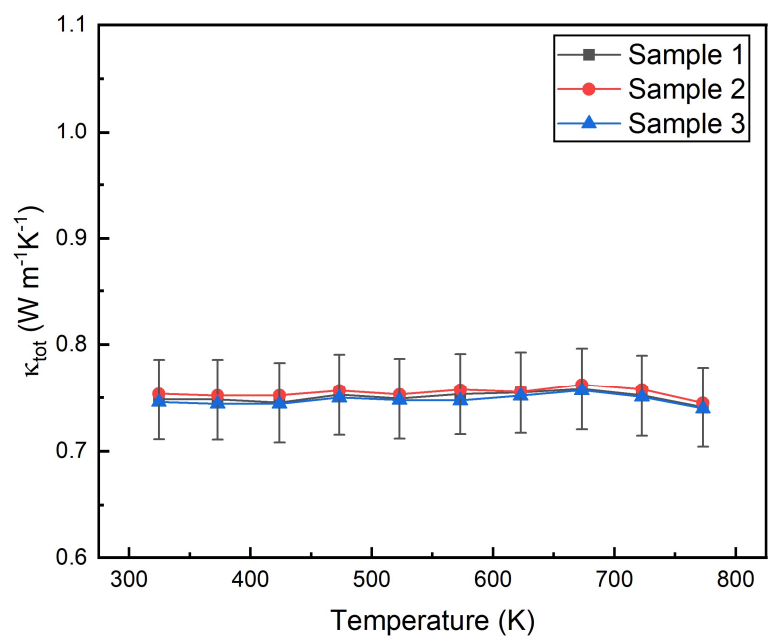

**Figure S12.** Thermal conductivity of $Mg_{3.2}Sb_{1.47}Bi_{0.5}Te_{0.02}Ge_{0.01}$.

## 8. Phonon scattering of the Janus nanoparticles

The Callaway model [6] is a well-known theoretical basis to interpret the temperature dependent lattice thermal conductivity ($\kappa_{lat}$). Here, we will use the same basis to consider the effect of the Janus nanoprecipitates on the thermal conductivity according to the Matthiessen's rule.[7]

$$\kappa_{lat} = \frac{k_B}{2\pi^2 v} \int_0^{\frac{k_B \theta_D}{\hbar}} \tau_{tot} \left(\frac{\hbar \omega}{k_B T}\right)^2 \frac{\exp(\hbar \omega / k_B T)}{[\exp(\hbar \omega / k_B T) - 1]^2} \omega^2 d\omega \quad (S1)$$

$$\tau_{tot}^{-1} = \tau_U^{-1} + \tau_{PD}^{-1} + \tau_B^{-1} + \tau_{NP}^{-1} + \tau_{Janus}^{-1} \quad (S2)$$

$$\tau_{PD}^{-1} = A \cdot \omega^4 \quad (S3)$$

$$\tau_U^{-1} = BT \cdot \omega^2 \quad (S4)$$

$$\tau_B^{-1} = \frac{v}{l} \quad (S5)$$

where the frequency dependent relaxation time $\tau_{tot}$, $\tau_U$, $\tau_{ph}$, $\tau_B$, $\tau_{NP}$, and $\tau_{Janus}$, are the combined relaxation time, the partial contribution due to Umklapp phonon-phonon scattering, phonon-point defects, phonon-boundary, nanoparticles scattering, and phonon-Janus particles, respectively. The phonon scattering of the Janus particles could be considered as a combination of an average effect of the whole particle and partial contribution of the internal sub-nanoparticle, as follows:

$$\frac{1}{\tau_{jnp}} = \frac{1}{\tau_{\langle jnp \rangle}} + \sum_i \frac{1}{\tau_{np-i}} \quad (S6)$$

For the phonon scattering of nanoparticles embedded in an alloy, Majumdar et al.[8] have proposed a cross-section combination of the Rayleigh limit ($\sigma_l$) and Geometric limit ($\sigma_s$).

$$\frac{1}{\tau_{np}} = v\rho \left(\frac{1}{\sigma_s} + \frac{1}{\sigma_l}\right)^{-1} \quad (S7)$$

$$\sigma_s = 2\pi R^2 \tag{S8}$$

$$\sigma_l = \frac{4}{9}\pi R^2 \cdot \left(\frac{\Delta D}{D_0}\right)^2 \left(\frac{\omega R}{\upsilon}\right)^4 \tag{S9}$$

Where $\rho$ is the particles concentration, $\Delta D$ is the density difference between the nanoparticle and the matrix, $D_0$ is the density of the matrix, $\omega$ is the angular frequency, $v$ is the average phonon speed, and $R$ is the average radius of nanoparticles.

Considering our matrix is a heavily doped system with spherical nanoparticles, we only consider the Rayleigh limit ($\sigma_l$), the Eq. (S7-9) could be

$$\frac{1}{\tau_{np}} = \frac{\upsilon}{3}\frac{f_{np}}{R}\left(\frac{\Delta D}{D_0}\right)^2 \left(\frac{\omega}{\upsilon}R\right)^4 \tag{S10a}$$

Or considering both the Rayleigh limit ($\sigma_l$) and Geometric limit ($\sigma_s$).

$$\frac{1}{\tau_{np}} = \frac{\upsilon}{3}\frac{f_{np}}{R}\left[\frac{2}{9} + \frac{1}{\left(\frac{\Delta D}{D_0}\right)^2 \left(\frac{\omega}{\upsilon}R\right)^4}\right]^{-1} \tag{S10b}$$

Now we consider a general case, i.e. a given volume fraction of the Janus nanoparticle $f_{jnp}$ with an average particle size of $2R_{jnp}$, the average density of $D_{jnp}$, which contains $i$ sub-nanoparticles with an average particle size of $2R_{i\text{-snp}}$ and the average density of $D_{i\text{-snp}}$. Then the Eq. S6 could be further expressed as follows:

$$\frac{1}{\tau_{jnp}} = \frac{1}{\tau_{\langle jnp \rangle}} + \sum_i \frac{1}{\tau_{np-1}} \tag{S11}$$

$$\frac{1}{\tau_{jnp}} = \frac{\upsilon}{3}\frac{f_{\langle jnp \rangle}}{R_{\langle jnp \rangle}}\left(\frac{D_{\langle jnp \rangle} - D_0}{D_0}\right)^2 \left(\frac{\omega}{\upsilon}R_{\langle jnp \rangle}\right)^4 + \sum_i \frac{1}{3(m\upsilon)^3}\frac{f_{i-np}}{R_{i-np}}\left(\frac{D_{i-np} - D_{\langle jnp \rangle}}{D_{\langle jnp \rangle}}\right)^2 (\omega R_{i-np})^4$$

$$f_{\langle jnp \rangle} = \sum_i f_{i-np}$$

where a correction factor $m$ is introduced to interpret partial contribution of the internal sub-nanoparticle of Janus nanoparticle, and relative with average phonon speed $\upsilon$. Next, we input equation S11 back to S1 and S2, and introduce the parameter $x = \hbar\omega / k_B T$, and applied to our case.

$$\kappa_{lat} = \frac{k_B}{2\pi^2 \upsilon_s} \left(\frac{k_B T}{\hbar}\right)^3 \int_0^{k\theta_D} \tau_{tot} \frac{x^4 e^x}{(e^x - 1)^2} dx$$

$$\frac{1}{\tau_{tot}} = A' \cdot T^4 \cdot x^4 + B' \cdot T^3 \cdot x^2 + \frac{\upsilon}{l}$$

$$+ C' f_{\langle jnp \rangle} \left(\frac{D_{\langle jnp \rangle} - D_0}{D_0}\right)^2 T^4 R_{\langle jnp \rangle}^3 x^4 + \sum_i C' f_{i-np} \left(\frac{D_{i-np} - D_{\langle jnp \rangle}}{D_{\langle jnp \rangle}}\right)^2 T^4 R_{i-np}^3 x^4$$

$$A' = A \left(\frac{k_B}{\hbar}\right)^4$$

$$B' = B \left(\frac{k_B}{\hbar}\right)^2$$

$$C' = \frac{1}{3(m\upsilon)^3} \left(\frac{k_B}{\hbar}\right)^4$$

Or considering both the Rayleigh limit ($\sigma_l$) and Geometric limit ($\sigma_s$) in the $\tau_{\langle jnp \rangle}$

$$\frac{1}{\tau_{tot}} = A' \cdot T^4 \cdot x^4 + B' \cdot T^3 \cdot x^2 + \frac{\upsilon}{l}$$

$$+ C \frac{f_{\langle jnp \rangle}}{R_{\langle jnp \rangle}} \left( \frac{2}{9} \left( \frac{\hbar}{k_B} \right)^4 + \frac{1}{\left( \frac{D_{\langle jnp \rangle} - D_0}{D_0} \right)^2 T^4 R_{\langle jnp \rangle}^4 x^4} \right)^{-1} + \sum_i C' \frac{f_{i-np}}{R_{i-np}} \left( \frac{D_{i-np} - D_{\langle jnp \rangle}}{D_{\langle jnp \rangle}} \right)^2 T^4 R_{i-np}^4 x^4$$

$$A' = A \left( \frac{k_B}{\hbar} \right)^4$$

$$B' = B \left( \frac{k_B}{\hbar} \right)^2$$

$$C = \frac{1}{3\upsilon^3} \left( \frac{k_B}{\hbar} \right)^4$$

$$C' = \frac{1}{3(m\upsilon)^3} \left( \frac{k_B}{\hbar} \right)^4$$

**Table S2. The parameters used for theoretical calculations of the lattice thermal conductivity of Mg$_{3.2}$Sb$_{1.47}$Bi$_{0.5}$Te$_{0.02}$Ge$_{0.01}$ with Janus nanoprecipitates.** The experimental parameters are from the STEM and APT investigations.

| Parameter | Symbol | Value | Units |
|---|---|---|---|
| **Experiment parameter** | | | |
| Average phonon speed | | 1800 | m/s |
| Debye temperature | $\theta_D$ | 179.12 | K |
| Atoms per unit cell | $N$ | 5 | - |
| Unit cell volume | $V\_Janus$ | 1.34E-28 | m$^3$ |
| Lattice constants | $c\_Janus$ | 7.29 | Å |
| | $a\_Janus$ | 4.60 | Å |
| Volume fraction of Janus nanoparticle | $f_{jnp}$ | 3.45 | vol% |
| Average particle size of Janus nanoparticle | $2R_{jnp}$ | 1.48E-08 | m |
| Average density of Janus nanoparticle | $D_{jnp}$ | 7.46 | g/cm$^3$ |
| Matrix density | $D_0$ | 4.43 | g/cm$^3$ |
| Volume fraction of Bi-rich precipitate | $f_{Bi}$ | 1.65 | vol% |
| Average particle size of Bi-rich precipitate | $2R_{Bi}$ | 2.00E-08 | m |
| Average density of Bi-rich precipitate | $D_{Bi}$ | 9.80 | g/cm$^3$ |
| Volume fraction of Ge-rich precipitate | $f_{Ge}$ | 1.80 | vol% |
| Average particle size of Ge-rich precipitate | $2R_{Ge}$ | 1.00E-08 | m |
| Average density of Ge-rich precipitates | $D_{Ge}$ | 5.32 | g/cm$^3$ |
| **Constant parameter** | | | |
| Planck constant | $h$ | 6.63E-34 | Js |
| Boltzmann constant | $k_B$ | 1.38E-23 | J/K |
| **Fitting parameter** | | | |
| Point defect scattering coefficient | $A$ | 1.00E-40 | S$^2$ |
| Umklapp phonon scattering coefficient | $B$ | 1.05E-17 | s/K |
| Distance between Janus particles | $l$ | 5.00E-06 | m |
| Individual particle sound velocity adjustment coefficient | $n$ | 16 | - |